\documentclass[a4paper]{appolb}
\usepackage[english]{babel}
\usepackage{floatflt}
\usepackage[dvips]{color}
\usepackage{colortbl} 
\usepackage{fancybox} 
\usepackage{epsfig} 
\usepackage{multicol}
\usepackage{./LaTeX/styles/mcite}
\RequirePackage{rotating}
\RequirePackage{epsfig}
\RequirePackage[centertags,intlimits]{amsmath}
\RequirePackage{amssymb,amscd}
\RequirePackage{array}
\RequirePackage{./LaTeX/styles/multirow}
\RequirePackage{./LaTeX/styles/supertabular}
\RequirePackage{dcolumn}
\RequirePackage{xspace}
\RequirePackage{./LaTeX/styles/axodraw}
\RequirePackage{./LaTeX/styles/upgreek}
\RequirePackage{color}
\RequirePackage{./LaTeX/styles/colors}
\RequirePackage{calc}
\RequirePackage{ifthen}
\RequirePackage{./LaTeX/styles/mcite}
\RequirePackage{cite}

\begin{document}
\title{Searches for Flavour Changing Neutral Currents \\
and Lepton Flavour Violation at HERA
\footnote{Presented at 
the 10th International Conference on Supersymmetry and Unification of 
Fundamental Interactions (SUSY02), June 17-23, DESY Hamburg.}
}
\author{Dominik Dannheim\address{DESY and University of Hamburg \\ 
Notkestr. 85, D-22607 Hamburg, Germany \\
E-mail: dominik.dannheim@desy.de \\
(on behalf of the H1 and ZEUS collaborations) }
}
\maketitle
\abstract{
\begin{minipage}[c]{0.85\textwidth}
A summary of the recent searches for anomalous single top-quark production
and lepton flavour violation in high energy $e^\pm p$ collisions
is presented. 
Single top-quark production via an anomalous flavour changing neutral current
$\gamma-u-t$ coupling would lead to the reaction $ep\rightarrow etX$,
$t\rightarrow Wb$.
H1 observed an excess of events above the Standard Model prediction
in the leptonic decay channel of the $W$ boson, leading to an isolated lepton,
large missing transverse momentum and a jet with large transverse momentum. 
Five of these events are compatible with single top-quark production, compared to 
1.8 events expected from background processes. ZEUS observed no excess above the SM 
background expectation in the leptonic channel. 
Neither of the two experiments found a deviation from the SM background expectation
in the 3-jet final state, which would result from the hadronic decay of the $W$ boson.
Searches for lepton flavour violating interactions of the 
type $ep \rightarrow \mu X$ and $ep\rightarrow \tau X$ have been performed.
No evidence was found for lepton flavour violation and limits were derived
on the production of leptoquarks and $R_p$ violating squarks, which could 
mediate such interactions.
\end{minipage}
}

\section{Single top-quark production at HERA}
\label{sec-single-top-HERA}
\begin{figure}[h]
\begin{minipage}[c]{\textwidth}
\hskip0.5cm
\begin{minipage}[c]{0.45\textwidth}
\vskip0.3cm
\hskip1.5cm
\epsfig{figure = 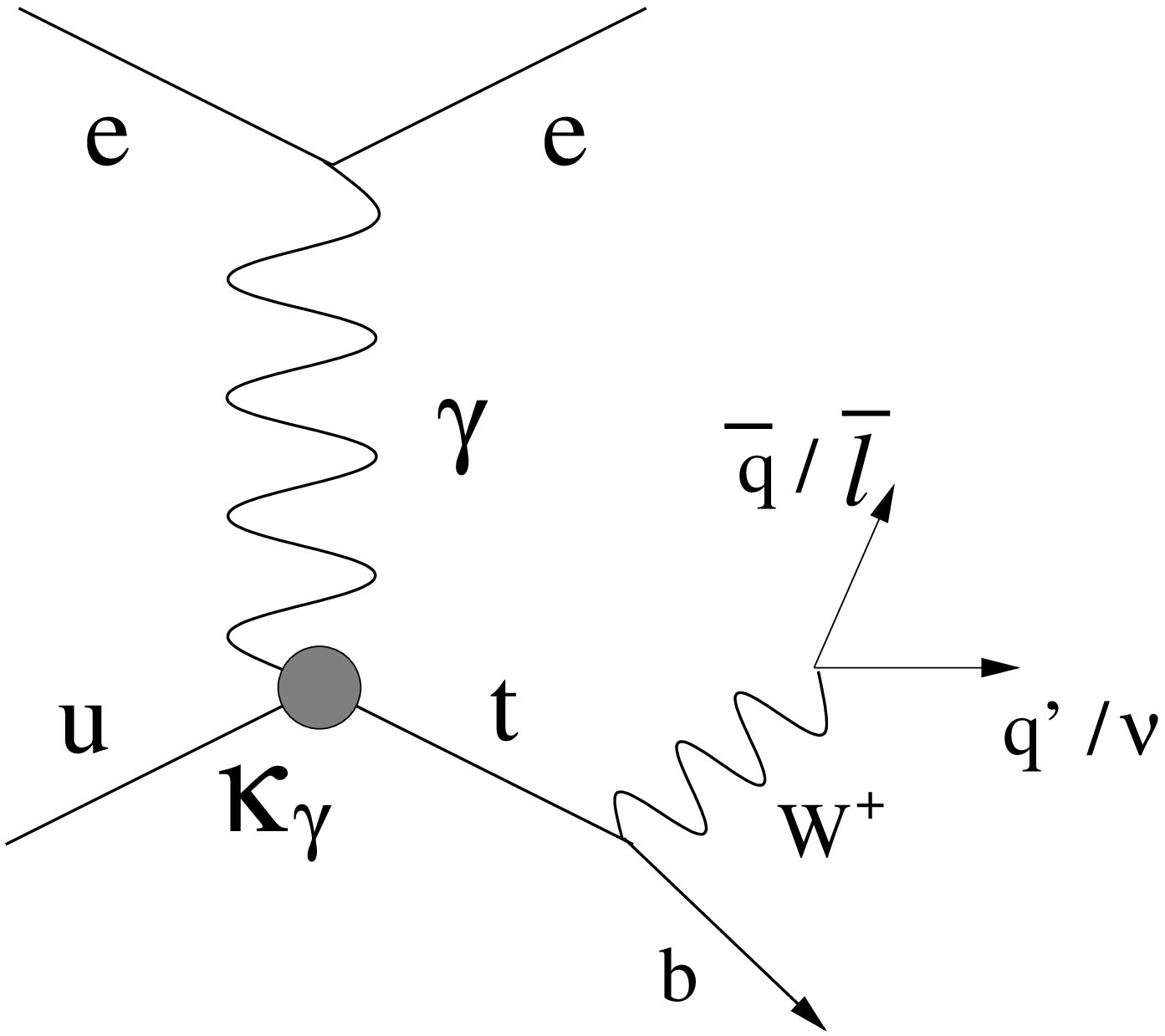,
width={4cm},angle=0, clip=}
\vskip0.35cm
\caption{Single top-quark production via flavour changing neutral
 current transitions at HERA.}
\label{fig-fcncdiagram}
\end{minipage}
\hskip0.5cm
\begin{minipage}[c]{0.45\textwidth}
\epsfig{figure = 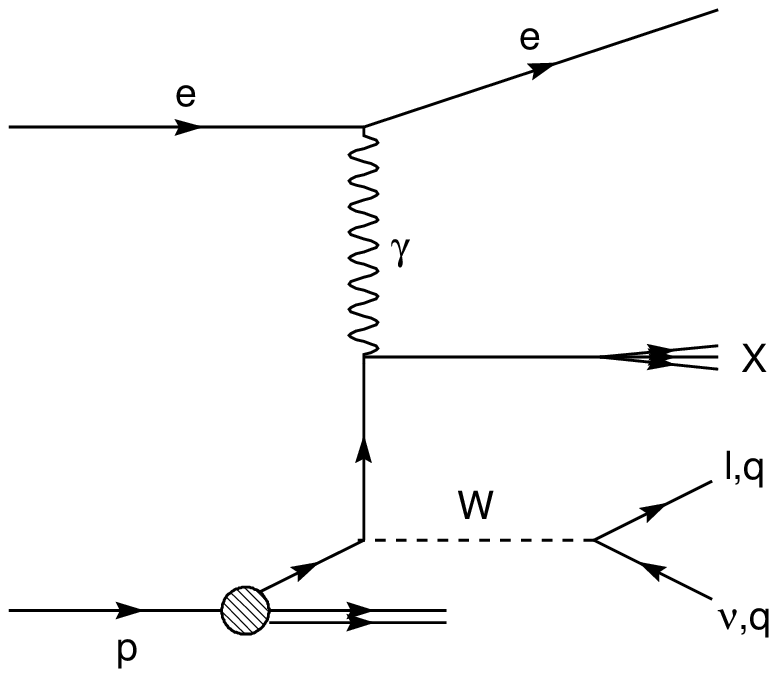,
width={5cm},angle=0, clip=}
\caption{Example for the production of real $W$ bosons at HERA.}
\label{fig-wdiagram}
\end{minipage}
\vfill
\end{minipage}
\end{figure}

Neutral current interactions in the Standard Model (SM) preserve the quark flavours in leading 
order perturbation theory. Quark flavour changing neutral current (FCNC) processes are
only present via higher order radiative corrections and are therefore strongly suppressed.
The SM cross section for single top-quark production at HERA is about 1~fb \cite{pr:d57:3040}.
Due to the large mass of the top quark, close to the electroweak symmetry breaking scale, 
possible deviations from the SM could be observed first in the top sector.
The production of single top quarks through FCNC is predicted by many extensions of the SM 
\cite{np:b454:527,pl:b426:393,pr:d58:073008,pr:d60:074015,pl:b457:186}.
The dominant contribution at HERA is expected to come from an anomalous coupling $\kappa_{tu\gamma}$
at the photon-u-quark-vertex (see Fig. \ref{fig-fcncdiagram}). 

The experimental signature of the 
process $ep\rightarrow etX$, $t\rightarrow Wb$ depends on the decay channel of the $W$ boson.
For leptonic decays, it consists of isolated leptons, missing transverse momentum and a hadronic
system at a large transverse momentum $p_T^X$. The main background for this process at HERA is
the production of real $W$ bosons (see Fig. \ref{fig-wdiagram}).
The hadronic decay of the $W$ boson leads to events
with one jet associated with the secondary $b$ quark and two jets from the secondary $W$ decay,
all of them at large transverse momenta. 

The observation of isolated lepton events with large values of $p_T^X$ by the H1 collaboration 
\cite{epj:c5:575} has motivated dedicated searches by the H1 and ZEUS collaborations for single 
top-quark production in both the leptonic 
and the hadronic channels \cite{misc:h1:eps02:1024,misc:zeus:eps01:650}. The results presented
here made use of all available HERA~I data.  

\subsection{Isolated lepton events}
\label{sect-isolept}
The preselection of isolated lepton events requires at least one isolated electron or muon 
with high transverse momentum $p_T^l$, large missing transverse momentum $p_T^{miss}$ and at 
least one hadronic jet with transverse momentum $p_T^X$. H1 optimized the selection for
$W$-decay signatures \cite{misc:h1:eps02:1024}. ZEUS performed an inclusive
analysis \cite{misc:zeus:eps01:650}, where the dominant background processes are neutral current 
deep inelastic scattering events (electron channel) and Bethe-Heitler muon pair production 
(muon channel). 

The expectations for direct $W$ production were simulated in leading order QCD calculations
by the generator EPVEC \cite{NP:B375:3}. The QCD NLO corrections change the $W$ boson production 
cross section by less than 10\% at large values of $p_T^X$ \cite{hep-ph/0203269}.

\begin{figure}
\hskip0.5cm
\begin{minipage}[c]{0.93\textwidth}
\begin{minipage}[c]{0.45\textwidth}
\epsfig{figure = 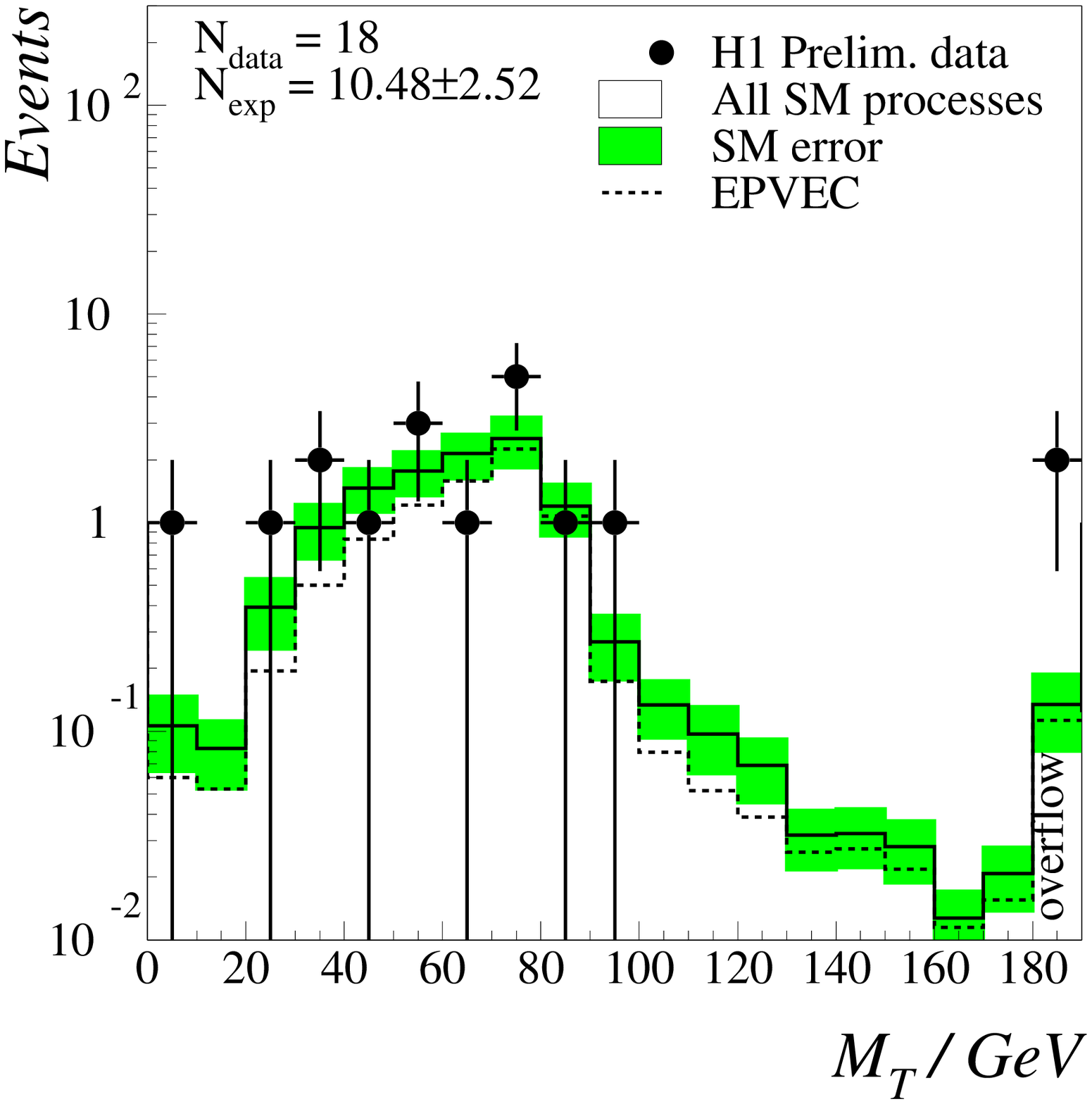,
width={\textwidth}, angle=0, clip=}
\end{minipage}
\hskip0.5cm
\begin{minipage}[c]{0.45\textwidth}
\epsfig{figure = 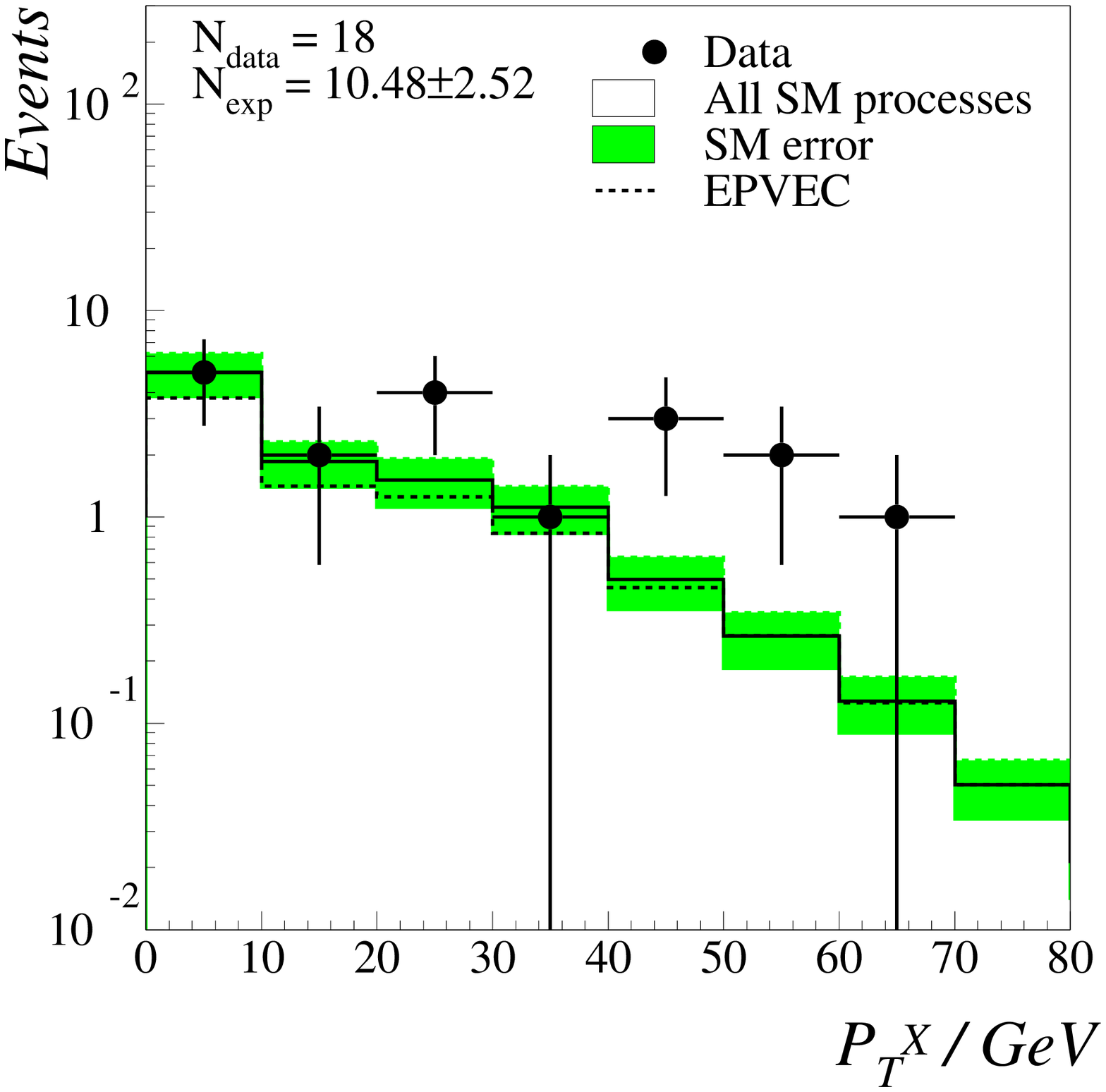,
width={\textwidth}, angle=0, clip=}
\end{minipage}
\caption{Distribution of the transverse mass (left) and the hadronic transverse momentum (right) for
the H1 isolated lepton events, compared to the SM prediction. Only $e^+p$ data are shown.}
\label{fig-h1-isolept}
\end{minipage}
\end{figure}

The H1 experiment observed 18 events in the data, while 10.5 events are expected from SM 
processes.
Figure \ref{fig-h1-isolept} (left) shows the distributions of
the H1 candidate events in the lepton-neutrino transverse mass $M_T$ ($P_T^{miss}$ being
attributed to the hypothetical neutrino). A Jacobian peak around the $W$ boson
mass is observed, as expected from the production of real $W$ bosons. 
Figure \ref{fig-h1-isolept} (right)
shows the distribution of the events in $p_T^X$. At low values of $p_T^X$,
the measured rates agree with the SM expectation, while at large values of $p_T^X$, 
more events are found than expected: Integrated over all $p_T^X$, 18 isolated lepton 
events were observed, while 10.5 events are expected from SM processes. For 
$p_T^X>40$ GeV, six events are observed with only one expected.

\begin{figure}
\hskip0.5cm
\begin{minipage}[c]{0.93\textwidth}
\begin{center}
\epsfig{figure = 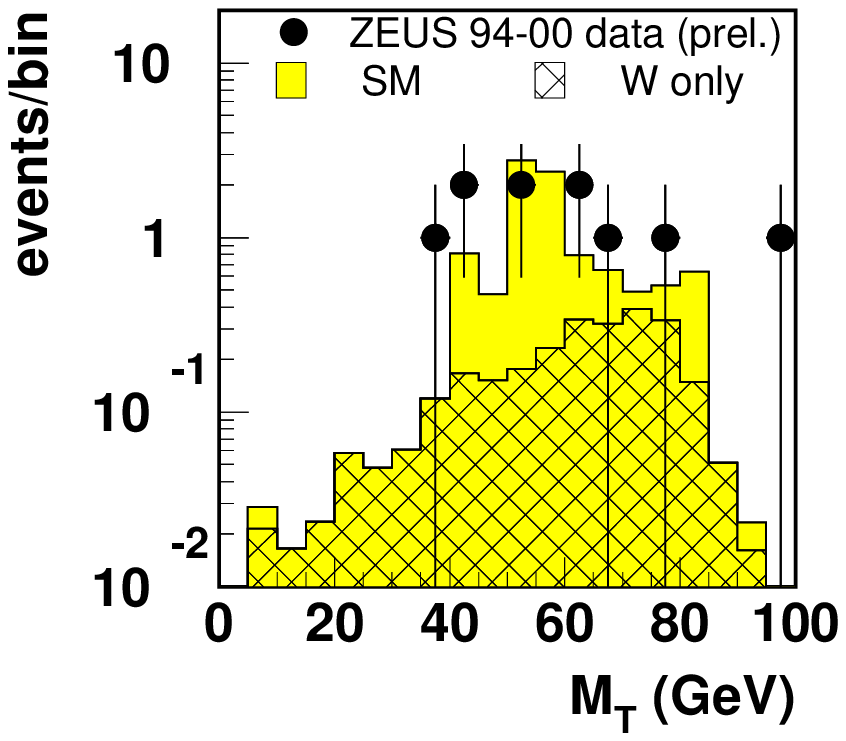,
width={6cm}, angle=0, clip=}
\epsfig{figure = 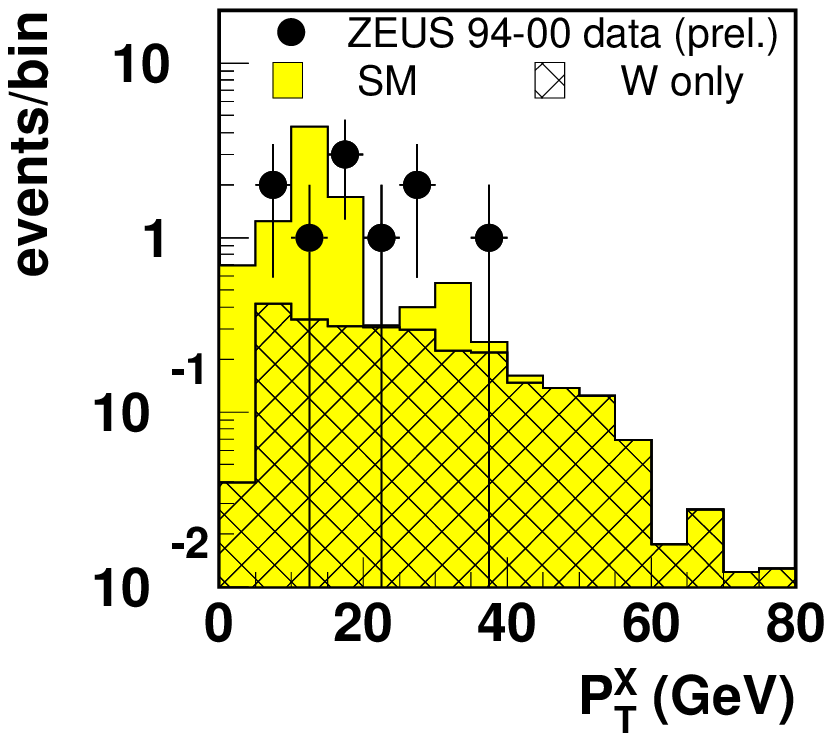,
width={6cm}, angle=0, clip=}
\caption{Distribution of the transverse mass of the lepton-neutrino system (left) and
of the hadronic transverse momentum (right) for the ZEUS isolated electron events,
compared to the SM prediction.}
\label{fig-zeus-isolept-el}
\end{center}
\begin{center}
\epsfig{figure = 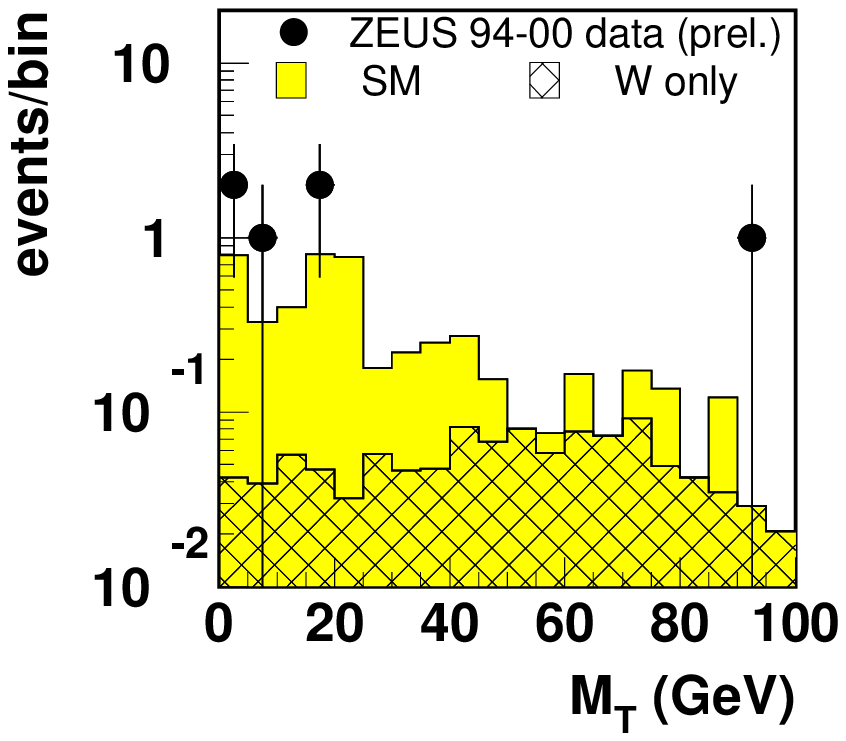,
width={6cm}, angle=0, clip=}
\epsfig{figure = 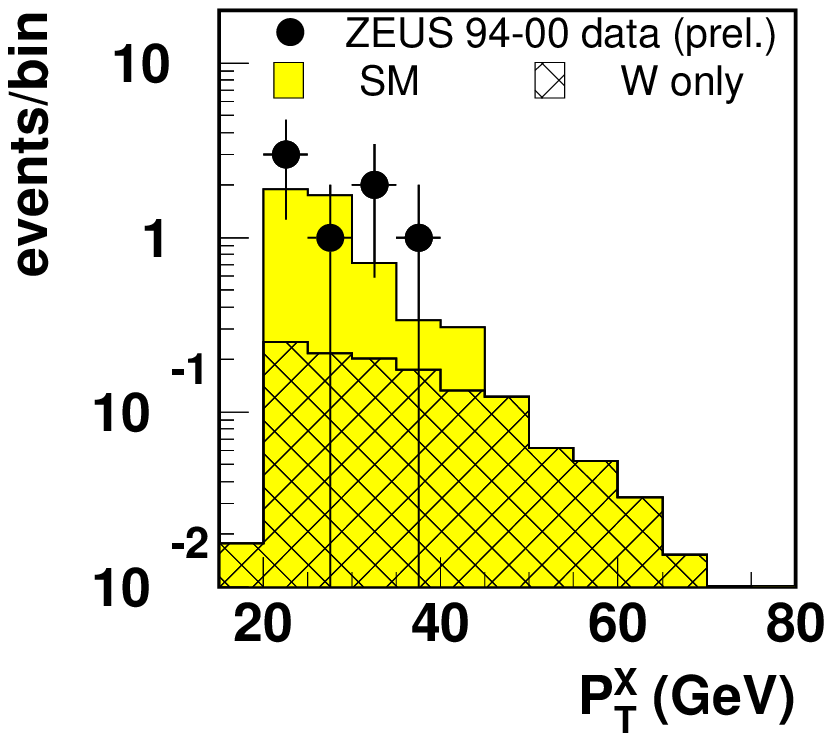,
width={6cm}, angle=0, clip=}
\caption{Distribution of the transverse mass of the lepton-neutrino system (left) and
of the hadronic transverse momentum (right) for the ZEUS isolated muon events,
compared to the SM prediction.}
\label{fig-zeus-isolept-mu}
\end{center}
\end{minipage}
\end{figure}

ZEUS finds agreement with the SM, both in the rate and in the distribution of the observed events.
Figures \ref{fig-zeus-isolept-el} and \ref{fig-zeus-isolept-mu} show the distributions of
isolated electron and muon events in $M_T$ and $p_T^X$. 

Table \ref{tab-isolept} compares the number of isolated lepton events observed at HERA with
those expected.
\begin{table}
\hskip0.5cm
\begin{minipage}[c]{0.93\textwidth}
\begin{center}
\begin{tabular}{|c||c|c|c|} \hline
 & H1 preliminary & H1 preliminary & ZEUS preliminary \\ 
 &  $e^+ p$, $102~\mathrm{pb}^{-1}$ &  $e^- p$, $14~\mathrm{pb}^{-1}$ & $e^\pm p$, $130~\mathrm{pb}^{-1}$ \\ \hline
$p_T^X$ range & Obs./expected & Obs./expected & Obs./expected \\ \hline \hline
$p_T^X>0~\mathrm{GeV} $  & 18 / 10.5 $\pm$ 2.5 & 0 / 1.8 $\pm$ 0.4 & 17 / 16.4 $\pm$ 1.7 \\ \hline  
$p_T^X>25~\mathrm{GeV} $ & 10 / 2.8 $\pm$ 0.7 & -                 &  2 / 2.4 $\pm$ 0.2 \\ \hline  
$p_T^X>40~\mathrm{GeV} $ & 6 / 1.0 $\pm$ 0.3 & -                 &  0 / 1.0 $\pm$ 0.1 \\ \hline  
\end{tabular}
\end{center}
\caption{Number of observed lepton events, compared to the SM background expectations.
The results from ZEUS for $p_T^X>25~\mathrm{GeV}$ and $p_T^X>40~\mathrm{GeV}$ are
with additional single top selection cuts applied, as described in section \ref{singletoplept}.}
\label{tab-isolept}
\end{minipage}
\end{table}
\subsection{Single top search in the leptonic channel}
\label{singletoplept}
The selections were further optimized for a signal from 
single top-quark production by requiring a large value of $p_T^X>40$ GeV
(ZEUS) and $>35$ GeV or $>25$ GeV (H1, depending on the hadronic jet angular range).
The $W$ component in the ZEUS selection was enhanced by applying additional 
cuts on the energy-momentum balance in the calorimeter and on the missing
transverse momentum, after correcting for the muon-track momentum.
Muon-type events were also rejected, if a second muon was found in the
event. H1 required a positive lepton charge and a value of $M_T$ above
10 GeV in order to reduce the contribution from processes with off-shell $W$ bosons.
After these cuts, ZEUS observed no event, while 1.0 is
expected from SM backgrounds, which are dominated by $W$ production,
whereas H1 observed 5 events, while $1.8\pm0.5$ are expected.

\begin{figure}
\hskip0.5cm
\begin{minipage}[c]{0.93\textwidth}
\vfill
\begin{center}
\epsfig{figure = 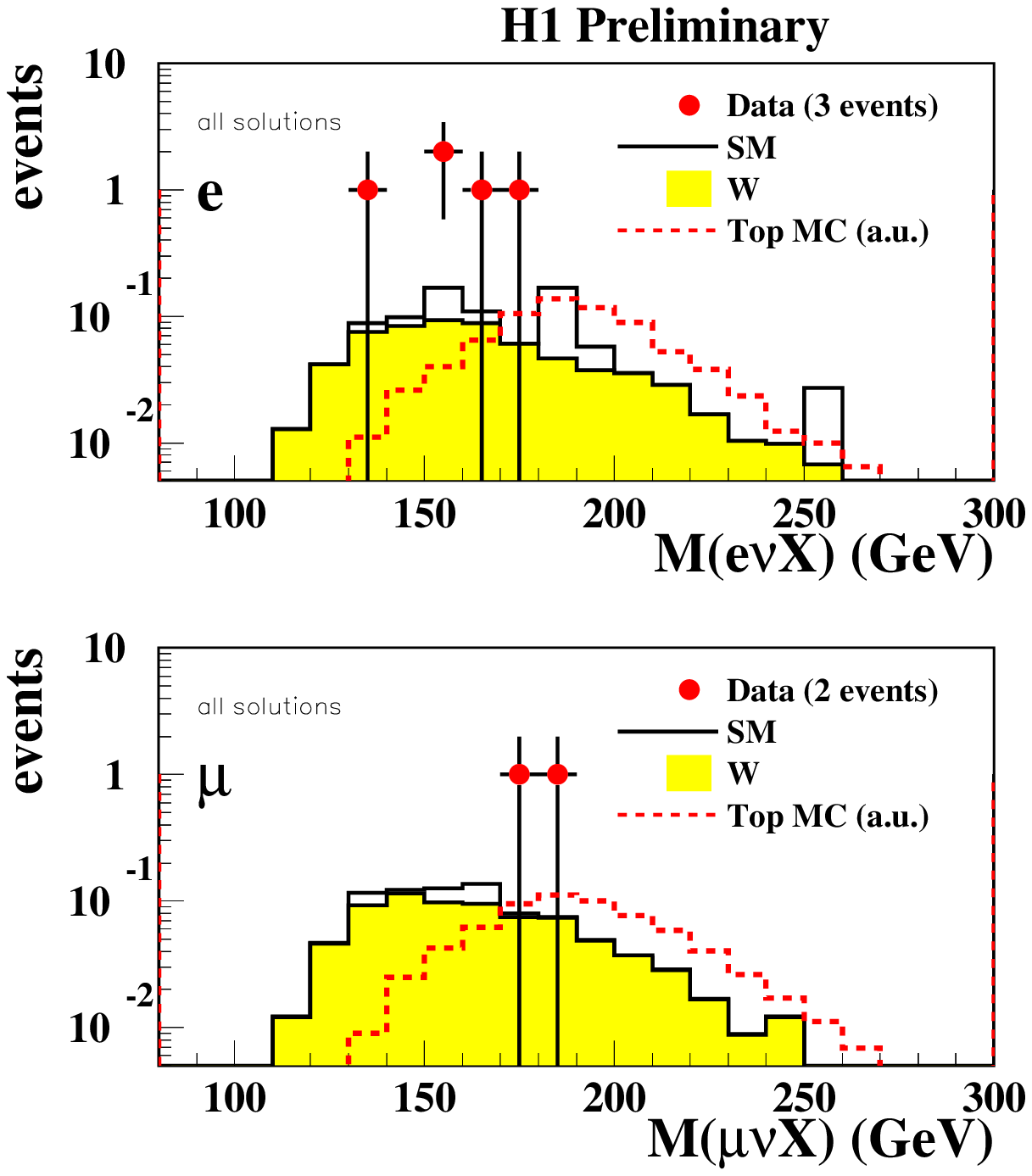,
width={8cm}, angle=0, clip=}
\end{center}
\caption{Distributions of the reconstructed invariant mass of the lepton-neutrino-hadron 
system for the H1 top candidates, compared to expectations from SM $W$ boson
production and a hypothetical single top-quark production signal (normalized to one event).}
\label{fig-h1-fcnc-lept-finalcand}
\vfill
\end{minipage}
\end{figure}

Assigning the missing transverse momentum to a hypothetical neutrino and assuming
the lepton-neutrino-system to originate from a $W$ boson decay, the 
lepton-neutrino-hadron invariant mass $M_{l\nu X}$ of the H1 candidates can be computed.
The neutrino kinematics were obtained either from the W mass constraint or
from the energy-momentum-balance, if the scattered lepton was observed in
the calorimeter. Two mass solutions were obtained for each event. For three
of the candidate events only one solution is compatible with the kinematic 
constraints, while for the other two candidates both solutions are acceptable. 
The reconstructed mass is compatible with that of the top quark for most of the events,
as shown in Fig. \ref{fig-h1-fcnc-lept-finalcand}.

\subsection{Single top search in the hadronic channel}

The hadronic decay of the $W$ boson results in events with one high-$p_T$ jet 
associated with the secondary $b$ quark and two high-$p_T$ jets from the 
secondary $W$ decay. The event selection required three jets with $p_T^{jet1}>40$ GeV,
$p_T^{jet2}>25$ GeV and $p_T^{jet3}>14$ GeV (ZEUS) or $p_T^{jet1}>40$ GeV,
$p_T^{jet2}>25$ GeV and $p_T^{jet3}>20$ GeV (H1). In both cases, the photoproduction
background was reduced by requiring one of the 2-jet masses and the 3-jet mass
to be compatible with the $W$ boson mass and the top mass, respectively. 

\begin{figure}
\hskip0.5cm
\begin{minipage}[c]{0.93\textwidth}
\begin{minipage}[c]{0.45\textwidth}
\epsfig{figure = 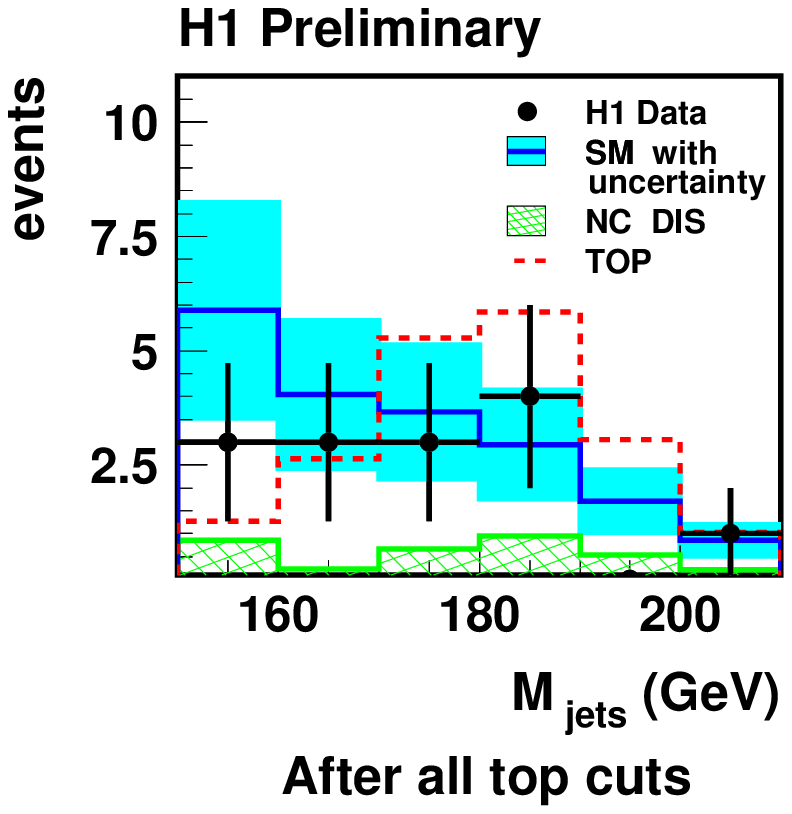,
width={6cm}, angle=0, clip=}
\end{minipage}
\hskip1cm
\begin{minipage}[c]{0.45\textwidth}
\epsfig{figure = 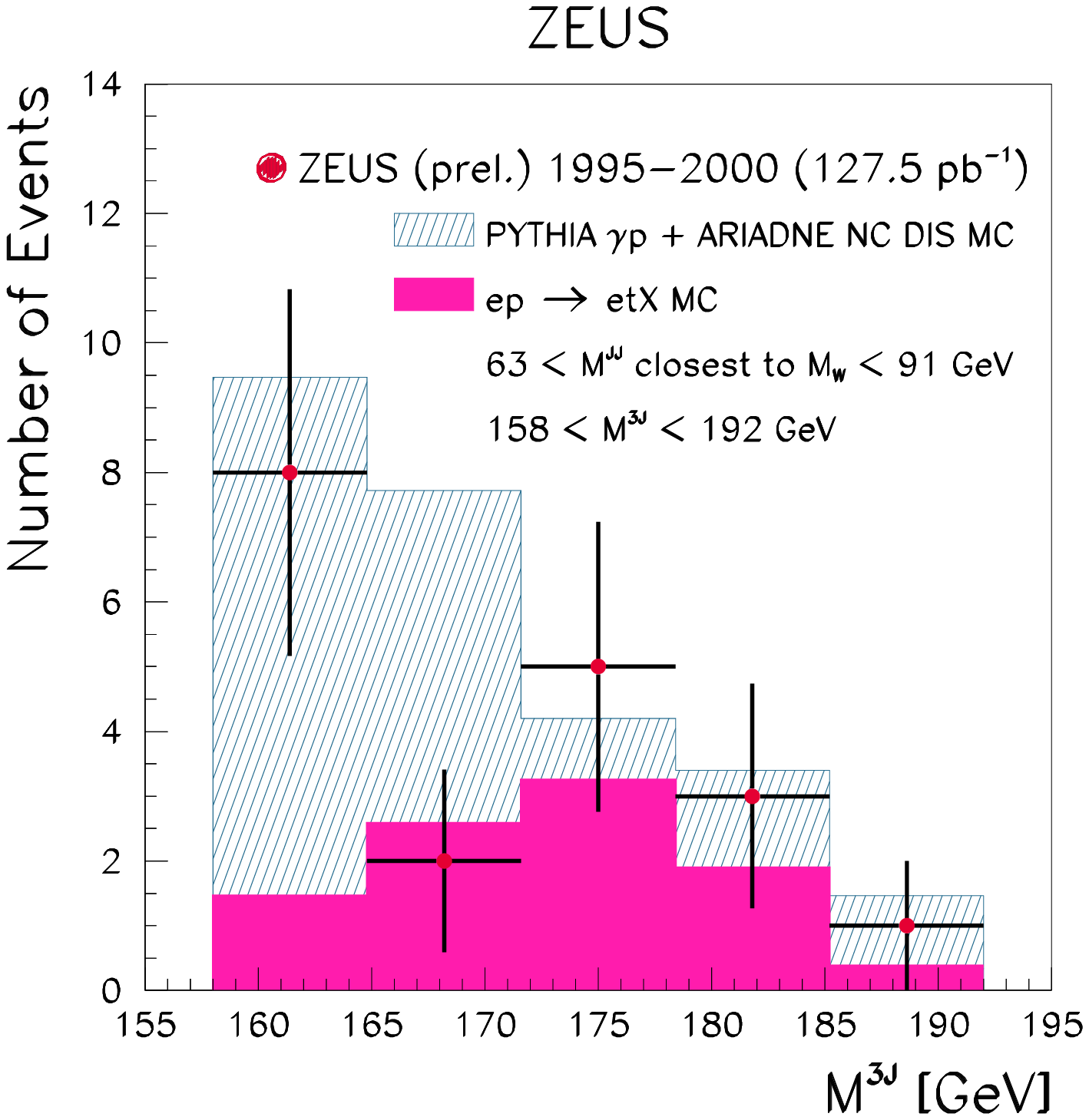,
width={6cm}, angle=0, clip=}
\end{minipage}
\caption{Distribution of the 3-jet mass of the H1 single top
candidates (left) and the ZEUS single top candidates (right)
in the hadronic channel, compared to expectations from SM photoproduction
background and a hypothetical signal from single top-quark production
with arbitrary normalisation.}
\label{fig-fcnc-m3jets}
\end{minipage}
\end{figure}

Figure \ref{fig-fcnc-m3jets} shows the 
distribution of the 3-jet mass in the selected mass windows for the
H1 and ZEUS candidate events. The normalisation uncertainties of the photoproduction
background, which was only evaluated in leading order of QCD, were reduced by 
normalising the simulated event rates 
to those observed in the low-mass domain. With the selection cuts described above,
H1 observed 14 events, while 19.6 $\pm$ 7.8 are expected, whereas ZEUS observed
19 events and expects 20.0. 

In both experiments the number of observed events in the hadronic channel is in
agreement with SM expectations.
The 95\% C.L. upper limit on the single top-quark production cross
section from the hadronic channel of the H1 analysis corresponds to an upper 
limit of 5.4 events from single top quark decays in the electron and muon channels. 
The interpretation of the observed lepton events as anomalous top-quark production 
is therefore not ruled out by the H1 analysis in the hadronic channel.

\subsection{Limits on FCNC couplings}

Upper limits on single top-quark production cross sections were set by both 
experiments, combining the leptonic and hadronic channels. Using an effective
Lagrangian, which describes the FCNC top-quark interactions \cite{pr:d60:074015}, 
they were converted into
95\% C.L. upper bounds on the anomalous coupling $\kappa _{tu\gamma}$ of 0.19 (ZEUS)
and 0.22 (H1). The ZEUS 95\% C.L. upper limit on the top cross section corresponds to 
an expectation of at most two top events in the H1 leptonic channel analysis.

\begin{figure}
\hskip0.5cm
\begin{minipage}[c]{0.93\textwidth}
\vfill
\begin{center}
\epsfig{figure = 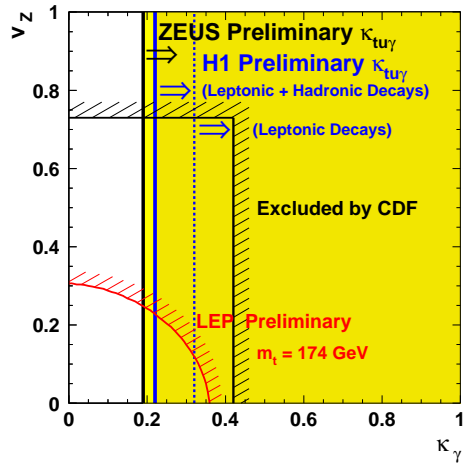,
width={6cm}, angle=0, clip=}
\end{center}
\caption{Limits from LEP, TEVATRON and HERA on anomalous couplings of the top quark 
to other quarks and neutral gauge bosons. The limits from H1 and ZEUS apply to
the coupling $\kappa _{\gamma,u}$ to the $u$ quark only. Both the H1 limit for
leptonic decays and the combined limit for leptonic and hadronic decays of the top
quark are indicated, whereas for ZEUS only the combined limit is indicated.}
\label{fig-h1-fcnc-limits}
\vfill
\end{minipage}
\end{figure}

The LEP \cite{pl:b494:33-tmp-3d28599b} and TEVATRON \cite{prl:80:2525} experiments have performed similar searches for single top
production and rare top decays, respectively. They are sensitive to anomalous
top-quark couplings to $u$ and $c$ quarks through both the photon and the $Z$ boson.
The results of these searches are compared to the HERA limits in 
Fig. \ref{fig-h1-fcnc-limits}. 
It is evident that HERA is competitive in searches for FCNC couplings
through photon interactions.

\section{Lepton-flavour violation}
Lepton flavour is conserved within the SM. The observation of neutrino oscillations \cite{prl:81:1562,*prl:87:071301},
however, is a hint that lepton flavour might be violated for the charged leptons. Minimal extensions of the SM 
\cite{ptp:28:870,*arevns:49:481}, that
allow for finite neutrino masses, do not predict measurable rates of LFV at current collider
experiments. Many extensions of the SM, however, involve LFV interactions.
Such models include grand unified theories \cite{pr:d10:275,*prl:32:438,*prep:72:185}, models based on supersymmetry 
\cite{prep:110:1,*prep:117:75},
compositeness \cite{pl:b153:101,*pl:b167:337} and technicolour 
\cite{np:b155:237,*np:b168:69,*pr:d20:3404,*prep:74:277}.

\begin{figure}
\hskip0.5cm
\begin{minipage}[c]{0.93\textwidth}
\begin{minipage}[c]{0.45\textwidth}
\epsfig{figure = 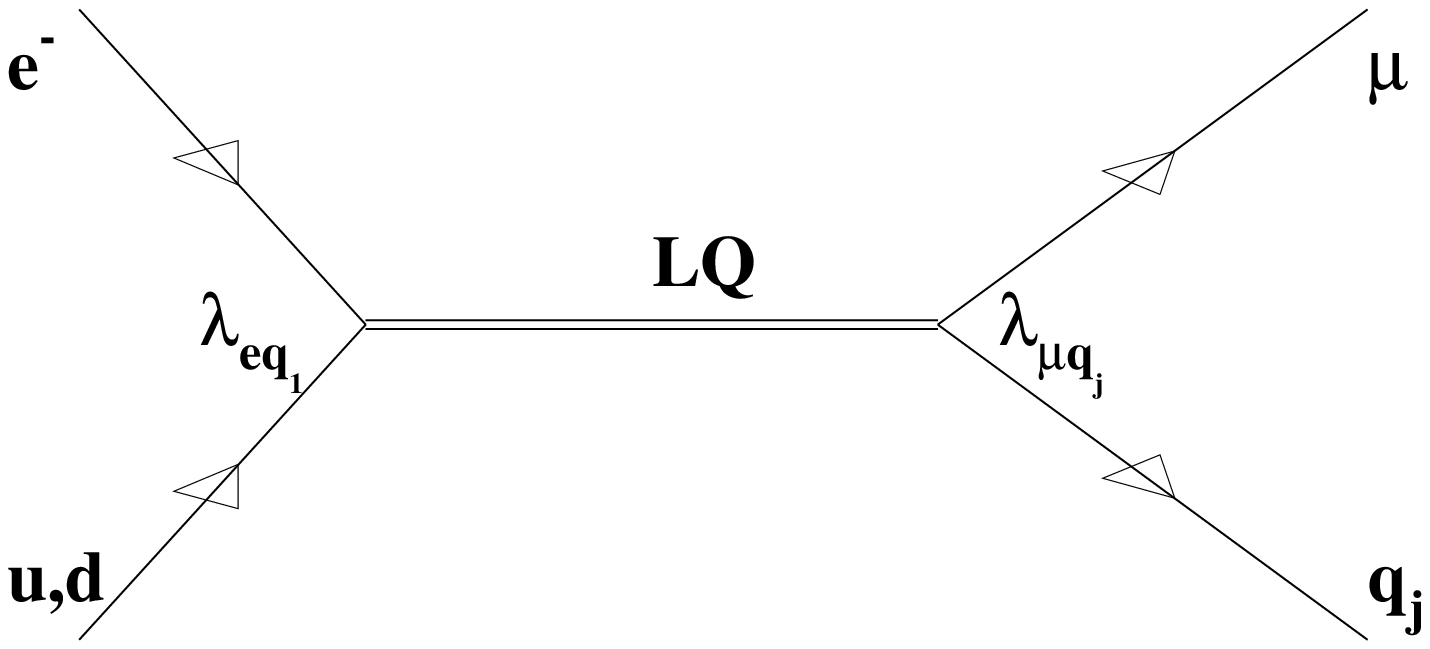,
width={6cm}, angle=0, clip=}
\end{minipage}
\hskip0.5cm
\begin{minipage}[c]{0.45\textwidth}
\epsfig{figure = 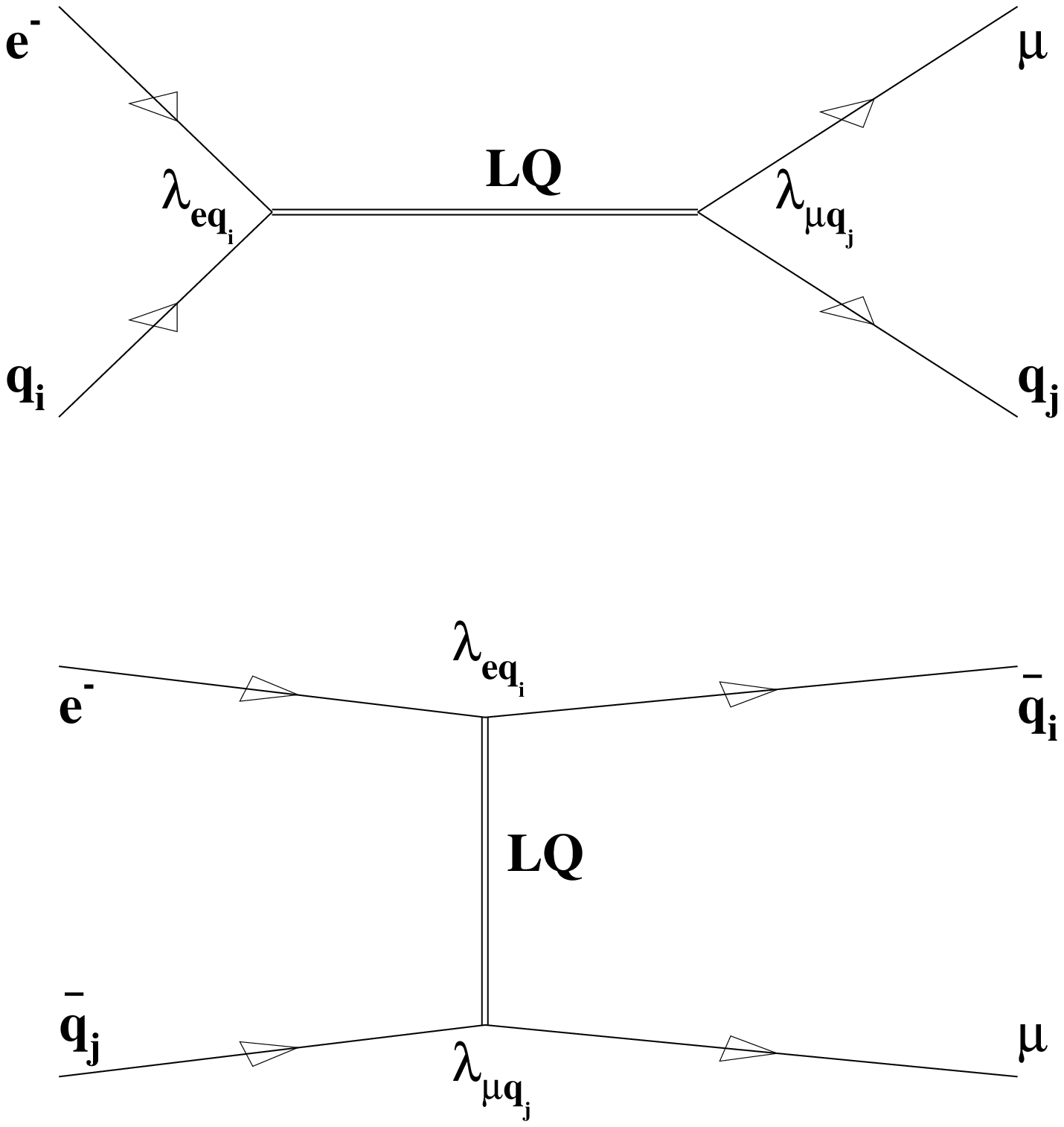,
width={6cm}, angle=0, clip=}
\end{minipage}
\caption{Example Feynman diagrams for LFV through LQs at HERA. For low mass LQs the dominant contribution is due to an $s$-channel
resonance (left). For high mass LQs both $u$- and $s$-channels contribute (right).}
\label{fig-zeus-lfv-diagram}
\end{minipage}
\end{figure}

At HERA, LFV can be detected in reactions of the type $ep\rightarrow lX$, where
$l$ is either a $\mu$ or a $\tau$ and $X$ denotes the hadronic final state. Figure \ref{fig-zeus-lfv-diagram}
shows examples for such reactions, where leptoquarks (LQs) mediate the LFV reaction.
LQs are bosons that carry
both lepton (L) and baryon (B) numbers and have lepton-quark Yukawa couplings. A LQ that couples to leptons of
two different generations would induce LFV. The Buchm\"uller-R\"uckl-Wyler (BRW) model \cite{pl:b191:442} was used to classify
LQ species and to evaluate cross sections for the LQ-induced processes. A total of 14 LQ types are allowed, seven scalars and
seven vectors, with fermion number $F=L+3B=0,2$. For a LQ mass $M_{LQ}$ below the HERA centre of mass energy 
($\sqrt{s}=300~\mathrm{GeV}$ for 1994-1997 data, $\sqrt{s}=318~\mathrm{GeV}$ for 1998-2000 data), it can be 
produced in a narrow resonance, with the $s$ channel giving
the dominant contribution (Fig. \ref{fig-zeus-lfv-diagram} left). For a LQ mass above the HERA centre of mass energy, 
both the $s$ and $u$ channel contribute (Fig. \ref{fig-zeus-lfv-diagram} right). In this case the reaction can
be approximated by a contact interaction and the cross section becomes proportional to 
$(\frac{\lambda_{eq_i}\lambda_{lq_j}}{M_{LQ}^2})^2$, where $\lambda_{eq_i}$ and $\lambda_{lq_j}$ represent the 
couplings as shown in Fig. \ref{fig-zeus-lfv-diagram}. 

The LFV processes can also be mediated by
$R$-parity-violating SUSY particles. 

Indirect searches for LFV processes \cite{zfp:c61:613} have yielded strong constraints
for cases where light quarks are involved. However, in some cases involving heavy quarks, the sensitivity of HERA
extends beyond existing low-energy limits.

The H1 experiment has searched for LFV in $e^+p$ data from 1994-1997, corresponding to an integrated luminosity 
of 37 $~\mathrm{pb}^{-1}$ \cite{epj:c11:447}. The ZEUS experiment has searched for LFV of the type $e^+p\rightarrow \tau X$ in
data from 1994-1997, corresponding to an integrated luminosity of 47.7 $~\mathrm{pb}^{-1}$ and for LFV of the type
$e^\pm p\rightarrow \mu X$ in the full HERA I dataset from 1994-2000, corresponding to an integrated luminosity
of 130 $~\mathrm{pb}^{-1}$ \cite{pr:d65:92004,misc:zeus:eps02:906}. 

\subsection{Experimental signature for LFV}

The experimental signature for LFV at HERA is a high transverse momentum muon or tau lepton and a jet from
the struck quark. The lepton and the jet are in opposite directions in the azimuthal plane. 
The tau lepton further decays to an electron, a muon or hadrons and neutrino(s).
Hence the signature in this channel is an electron or muon at 
high transverse momentum $p_T^l$ or a narrow, pencil-like jet with low track multiplicity  
from the hadronic decay of the tau lepton. In all channels, the presence of a muon and/or neutrinos leads to
a large value of the missing transverse momentum, $p_T^{miss}$, measured in the calorimeter, in the direction 
of the final state lepton.
The background from SM processes is low and the detection efficiency is 40\% to 60\% in the muon
channel and 25\% to 30\% in the tau channel, depending on the LQ mass (for $M_{LQ}\ll \sqrt{s}$) and on the 
generation of the initial state quark (for $M_{LQ}\ll \sqrt{s}$). 

No candidate events for LFV were observed by either H1 or ZEUS. Therefore limits were set on
the production of LQs that mediate LFV interactions.

\subsection{Low mass LQ limits}

\begin{figure}
\hskip0.5cm
\begin{minipage}[c]{1.0\textwidth}
\begin{minipage}[c]{0.45\textwidth}
\epsfig{figure = 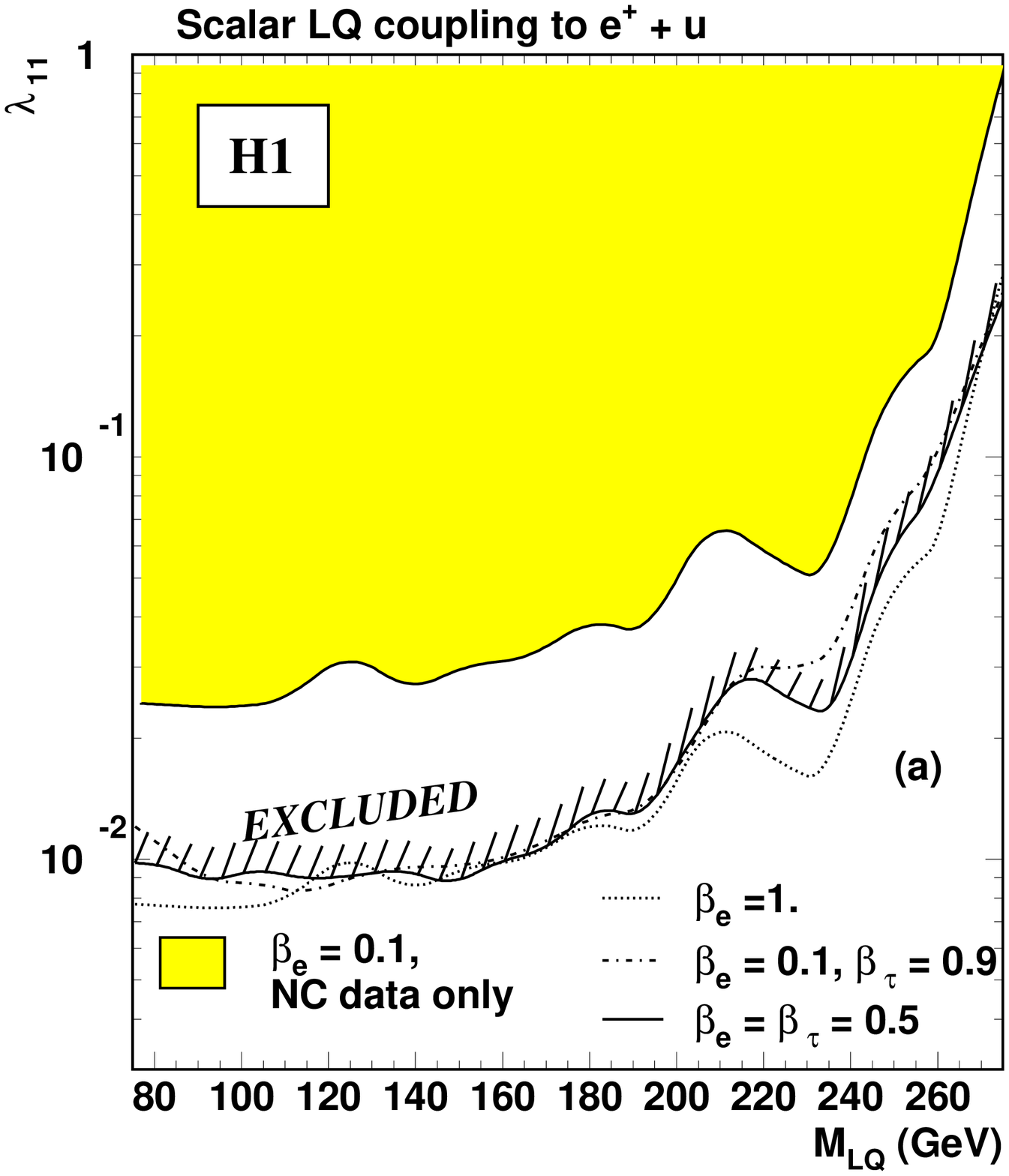,
width={6cm}, angle=0, clip=}
\caption{Result from H1 for the 95\% C.L. upper limit on the electron-valence-quark coupling of the $S_{1/2}^L$ LQ.
The limit is shown as a function of the LQ mass and for different branching ratios of the LQ decay into $e$ and $\tau$.}
\label{fig-h1-lfv-lowmass-tau}
\end{minipage}
\hskip0.6cm
\begin{minipage}[c]{0.45\textwidth}
\hskip0.3cm
\epsfig{figure = 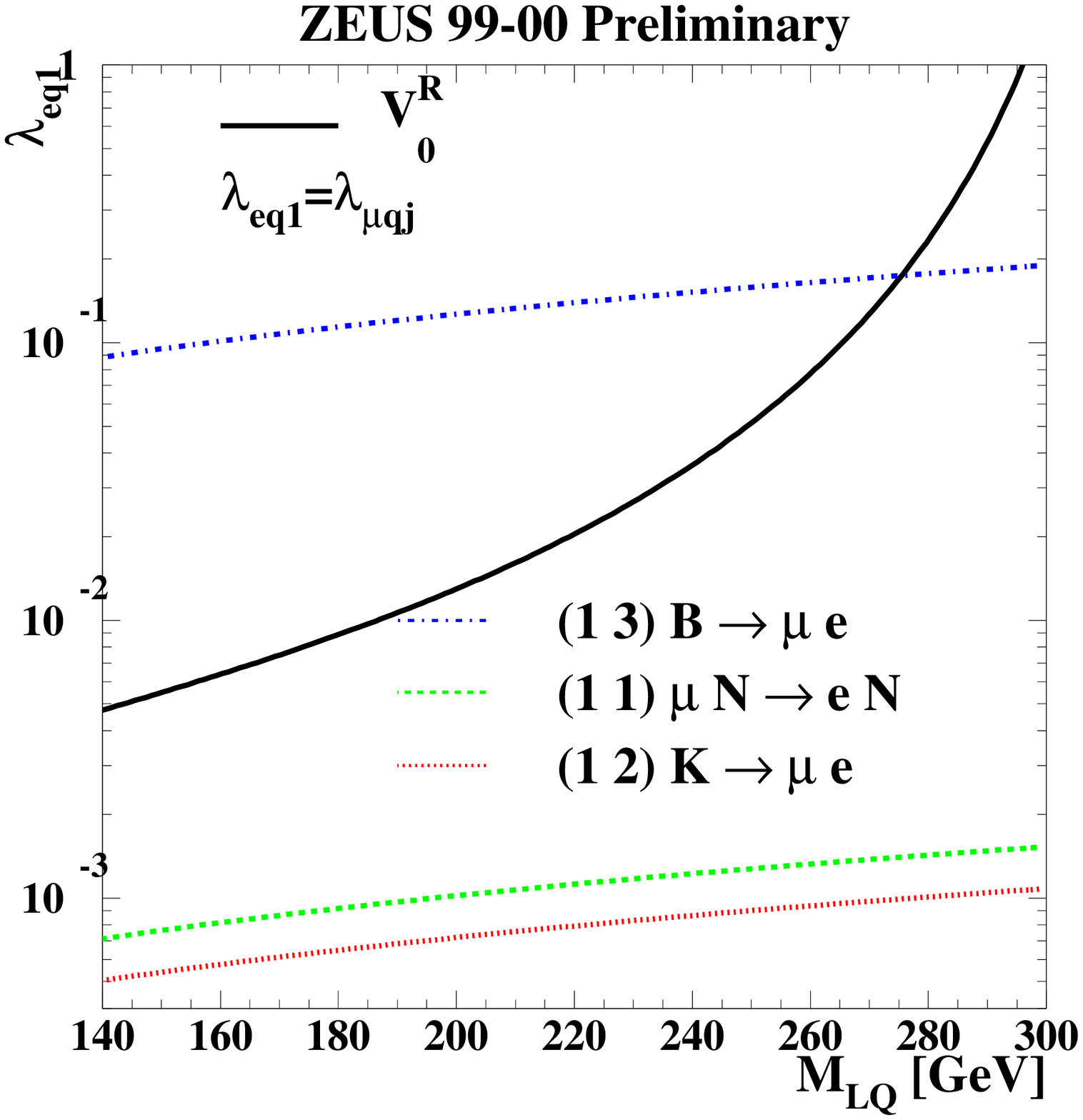,
width={6cm}, angle=0, clip=}
\caption{Result from ZEUS for the 95\% C.L. upper limit on the electron-valence-quark coupling of the $V_0^R$ LQ,
decaying into $\mu q_j$ (solid line), where $q_j$ denotes a down-type quark from any generation. The limit is shown as a 
function of the LQ mass. The dotted lines represent the limits from low energy experiments. The second number
in the parenthesis denotes the generation of $q_j$. }
\label{fig-zeus-lfv-mu-limits}
\end{minipage}
\end{minipage}
\end{figure}

Figure \ref{fig-h1-lfv-lowmass-tau} shows the 95\% C.L. limit on the coupling constant for a $S_{1/2}^L$ LQ
decaying into $\tau$, as obtained from the H1 analysis. The result is shown for different branching
ratios $\beta _l$ ($l=e,\tau$) of the LQ into $e$ and $\tau$.  Assuming an electromagnetic 
coupling strength and $\beta _\tau=0.9$, a limit of 275~GeV at 95\% C.L. has been set on the LQ mass.
The ZEUS collaboration obtained similar results in the $\tau$ channel. 

Figure \ref{fig-zeus-lfv-mu-limits} shows the 95\% C.L. limit on the coupling constant for a $V_0^R$ LQ, decaying
into $\mu$, as obtained from the ZEUS analysis. The result is compared to limits from low energy experiments.
The HERA limits are competitive to those obtained from rare meson decays in cases where heavy quarks are involved.
Assuming an electromagnetic coupling strength and $\beta _\mu = 0.5$, LQs with masses up to
301 $~\mathrm{GeV} $ are excluded at 95\% C.L., depending on the LQ species.
\subsection{High mass LQ limits}
Table \ref{tab-zeus-lfv-highmass-tau-limits-f2} shows the 95\% C.L. limits on
$(\frac{\lambda_{eq_i}\lambda_{lq_j}}{M_{LQ}^2})^2$ for the $\mu$ channel and $F=2$, as obtained by the
ZEUS collaboration. The limits improve those from rare $D$ meson decays, when quarks from the second generation
are involved. Similar results were obtained for $F=0$ and for the tau channel.
\input{colordvi}
\begin{table} [htbp]
\hskip0.5cm
\begin{minipage}[c]{0.93\textwidth}
\begin{center} 
\tiny
\begin{tabular}{|c||c|c|c|c|c|c|c|} \hline 
\multicolumn{4}{|c}{} & \multicolumn{4}{c|}{} \\ 
\multicolumn{2}{|c}{\footnotesize{$F=2$}} & \multicolumn{4}{c}{\Red{\small{Zeus Preliminary (94-00 combined limits)}}} & \multicolumn{2}{c|}{\tiny{\Blue{$\frac{\lambda_{eq_i}\lambda_{\mu q_j}}{M^2_{LQ}}$ (TeV$^{-2})$}} } \\ 
\multicolumn{4}{|c}{} & \multicolumn{4}{c|}{}\\ 
\hline 
\tiny{$q_i q_j$} & \tiny{$S_0^L$}   & \tiny{$S_0^R$}  & \tiny{$\tilde{S}_0^R$} & \tiny{$S_1^L$}  & \tiny{$V_{1/2}^L$} & \tiny{$V_{1/2}^R$} & \tiny{$\tilde{V}_{1/2}^L$} \\ 
   & $e^- u $    & $e^- u $   & $e^- d $   & $e^-(u+ \sqrt{2} d)$ & $e^- d $  & $e^-(u+d)$  & $e^- u $\\ 
   & $e^+\bar{u} $    & $e^+ \bar{u} $   & $e^+ \bar{d} $   & $e^+(\bar{u}+ \sqrt{2} \bar{d})$ & $e^+ \bar{d} $  & $e^+(\bar{u}+\bar{d})$  & $e^+\bar{u} $\\ \hline \hline 
   & $\mu N \rightarrow e N$  & $\mu N \rightarrow e N$ & $\mu N \rightarrow e N$ & $\mu N \rightarrow e N$ & $\mu N \rightarrow e N$ & $\mu N \rightarrow e N$ & $\mu N \rightarrow e N$ \\ 
1 1&\Blue{$7.6 \times 10^{-5} $}&\Blue{$7.6 \times 10^{-5}$}&\Blue{$7.6 \times 10^{-5}$}&\Blue{$2.3\times 10^{-5}$}&\Blue{$2.6 \times 10^{-5}$}&\Blue{$1.3 \times 10^{-5}$}&\Blue{$2.6 \times 10^{-5}$}\\ 
&\boldmath \RedViolet{$1.6$}&\boldmath \RedViolet{$1.6$}&\boldmath \RedViolet{$2.1$}  &\boldmath \RedViolet{$0.9$}&\boldmath \RedViolet{$0.9$}&\boldmath \RedViolet{$0.5$}&\boldmath \RedViolet{$0.6$}\\ \hline 
   & $K \rightarrow \pi \nu \bar{\nu}$ & $D \rightarrow \mu \bar{e}$ & $K \rightarrow \mu \bar{e}$ & $K \rightarrow \mu \bar{e}$ & $K \rightarrow \mu \bar{e}$ & $K \rightarrow \mu \bar{e}$ & $D \rightarrow \mu \bar{e}$ \\ 
1 2&\Blue{$10^{-3}$}&\Blue{$4$}&\Blue{$2.7 \times 10^{-5}$}&\Blue{$2.7 \times 10^{-5}$}&\Blue{$1.3 \times 10^{-5}$}&\Blue{$1.3 \times 10^{-5}$}&\Blue{$2$}\\ 
&\boldmath \RedViolet{$2.4$}&\Red{\doublebox{\Black{\boldmath{$2.4$}}}}& \boldmath \RedViolet{$2.6$}&\boldmath \RedViolet{$1.1$}&\boldmath \RedViolet{$1.6$}&\boldmath \RedViolet{$1.2$}&\Red{\doublebox{\Black{\boldmath{ $1.7$}}}}\\ \hline 
& $V_{ub}$&&$B \rightarrow \mu \bar{e}$&$V_{ub}$&$B \rightarrow \mu \bar{e}$&$B \rightarrow \mu \bar{e}$&\\ 
1 3&\Blue{$0.4$}&\boldmath \RedViolet{$*$}&\Blue{$0.8$}&\Blue{$0.4$}&\Blue{$0.4$}&\Blue{ $0.4$ }&\boldmath \RedViolet{$*$}\\ 
&\boldmath \RedViolet{$*$}&&\boldmath \RedViolet{$2.8$}&\boldmath \RedViolet{$1.4$}&\boldmath \RedViolet{$2.1$}&\boldmath \RedViolet{$2.1$}& \\ \hline 
&$K \rightarrow \pi \nu \bar{\nu}$ & $D \rightarrow \mu \bar{e}$ & $K \rightarrow \mu \bar{e}$ & $K \rightarrow \mu \bar{e}$ & $K \rightarrow \mu \bar{e}$ & $K \rightarrow \mu \bar{e}$ & $D \rightarrow \mu \bar{e}$ \\ 
2 1&\Blue{ $10^{-3}$ }&\Blue{ $4$}&\Blue{$2.7 \times 10^{-5}$}& \Blue{$2.7 \times 10^{-5}$}&\Blue{$1.3 \times 10^{-5}$}&\Blue{$1.3 \times 10^{-5}$}&\Blue{$2$}\\ 
&\boldmath \RedViolet{$2.1$}&\Red{\doublebox{\Black{\boldmath{$2.1$}}}} & \boldmath \RedViolet{$2.6$}&\boldmath \RedViolet{$1.1$}&\boldmath \RedViolet{$0.9$}&\boldmath \RedViolet{$0.5$}&\Red{\doublebox{\Black{\boldmath{$0.6$}}}}\\ \hline 
  & $\mu \rightarrow 3e$ & $\mu \rightarrow 3e$ & $\mu \rightarrow 3e$ & $\mu \rightarrow 3e$ & $\mu \rightarrow 3e$ & $\mu \rightarrow 3e$ & $\mu \rightarrow 3e$ \\ 
2 2&\Blue{$5\times 10^{-3}$}&\Blue{$5 \times 10^{-3}$}&\Blue{$1.6 \times 10^{-2}$}&\Blue{$1.3 \times 10^{-2}$}&\Blue{$8 \times 10^{-3}$}&\Blue{$3.7 \times 10^{-3}$}&\Blue{ $2.5 \times 10^{-3}$}\\ 
  & \boldmath \RedViolet{$5.7$} & \boldmath \RedViolet{$5.7$} & \boldmath \RedViolet{$3.8$} & \boldmath \RedViolet{$1.8$}  & \boldmath \RedViolet{$1.9$} & \boldmath \RedViolet{$1.6$ } & \boldmath \RedViolet{$2.9$ } \\ \hline 
  & $B \rightarrow l \nu X$ & & $B \rightarrow \bar{\mu} e K$ & $B \rightarrow \bar{\mu} e K$ & $B \rightarrow \bar{\mu} e K$ & $B \rightarrow \bar{\mu} e K$ & \\ 
2 3&\Blue{$4$}&\boldmath \RedViolet{$*$}&\Blue{$0.6$}&\Blue{$0.3$}&\Blue{$0.3$}&\Blue{$0.3$}&\boldmath \RedViolet{$*$}\\ 
&\boldmath \RedViolet{$*$ }&&\boldmath \RedViolet{$4.3$}&\boldmath \RedViolet{$2.1$}&\boldmath \RedViolet{$2.9$}&\boldmath \RedViolet{$2.9$}&\\\hline 
&$V_{ub}$& & $B \rightarrow \mu \bar{e}$ & $V_{ub}$ & $B \rightarrow \mu \bar{e}$ & $B \rightarrow \mu \bar{e}$ & \\ 
3 1&\Blue{$0.4$}&\boldmath \RedViolet{$*$}&\Blue{$0.8$}&\Blue{$0.4$}&\Blue{$0.4$}&\Blue{$0.4$}&\boldmath \RedViolet{$*$}\\ 
&\boldmath \RedViolet{$*$}&&\boldmath \RedViolet{$3.1$}&\boldmath \RedViolet{$1.5$}&\boldmath \RedViolet{$1.0$}&\boldmath \RedViolet{$1.0$}&\\ \hline 
  & $B \rightarrow l \nu X$ & & $B \rightarrow \bar{\mu} e K$ & $B \rightarrow \bar{\mu} e K$ & $B \rightarrow \bar{\mu} e K$ & $B \rightarrow \bar{\mu} e K$ & \\ 
 3 2 &\Blue{ $4$}   & \boldmath \RedViolet{$*$ } & \Blue{$0.6$ }   & \Blue{$0.3$}   & \Blue{$0.3$} & \Blue{$0.3$} & \boldmath \RedViolet{$*$ } \\ 
  & \boldmath \RedViolet{$*$ } & & \boldmath \RedViolet{$5.8$} & \boldmath \RedViolet{$2.9$ }  & \boldmath \RedViolet{$2.2$} & \boldmath \RedViolet{$2.2$ } & \\ \hline 
&&&$\mu \rightarrow 3e$ & $\mu \rightarrow 3e$ & $\mu \rightarrow 3e$ & $\mu \rightarrow 3e$ &  \\ 
3 3&\boldmath \RedViolet{$*$}& \boldmath \RedViolet{$*$}&\Blue{$1.6\times 10^{-2}$}&\Blue{$1.3 \times 10^{-2}$}&\Blue{$8 \times 10^{-3}$}&\Blue{$3.7 \times 10^{-3}$}&\boldmath \RedViolet{$*$}\\ 
&&&\boldmath \RedViolet{$7.6$}&\boldmath \RedViolet{$3.8$}&\boldmath \RedViolet{$4.0$}&\boldmath \RedViolet{$4.0$}&\\ \hline 
\end{tabular}
\caption{Result from ZEUS for 95\% C.L. upper limits on $(\frac{\lambda_{eq_i}\lambda_{lq_j}}{M_{LQ}^2})^2$
for $F=2$ LQs. $q_i$ and $q_j$ in the first column indicate the quark generations coupling to LQ-$e$ and LQ-$\mu$,
respectively. The low energy process, which provides the most stringent limit, and its value are given in the
first and second row of each column and compared to the ZEUS limit value in the third row of each column. The
ZEUS limits are enclosed in a box in cases where the ZEUS limits are lower than the ones from low energy experiments. 
The * indicates the cases where a top quark would have to be involved.}
\label{tab-zeus-lfv-highmass-tau-limits-f2}
\normalsize
\end{center}
\end{minipage}
\end{table} 
\section{Outlook}
The H1 and ZEUS experiments are expecting a much higher integrated luminosity within five years of operation from the 
HERA II running period. This will lead to a higher sensitivity for probing coupling 
strengths in searches for FCNC and LFV. It will in particular help to clarify the origin of the excess of isolated 
lepton events at high $p_T^X$, which were observed by the H1 collaboration.
In addition, both experiments have improved their detectors, in particular for the forward tracking. 
The enhanced capability for $b$-tagging will further improve the sensitivity for new phenomena such as anomalous FCNC 
and LFV. 

\section*{Acknowledgments}
I would like to thank all my colleagues from H1 and ZEUS for their outstanding work in producing these results.
I am grateful to T. Carli, M. Derrick, M. Kuze and P. Newman for carefully reading the manuscript.
\newpage
\def\bibname{\Large\bf References}
\def\refname{\Large\bf References}
\bibliographystyle{./l4z_default}
{\raggedright
\providecommand{\etal}{et al.\xspace}
\providecommand{\coll}{Coll.\xspace}
\catcode`\@=11
\def\@bibitem#1{%
\ifmc@bstsupport
  \mc@iftail{#1}%
    {;\newline\ignorespaces}%
    {\ifmc@first\else.\fi\orig@bibitem{#1}}
  \mc@firstfalse
\else
  \mc@iftail{#1}%
    {\ignorespaces}%
    {\orig@bibitem{#1}}%
\fi}%
\catcode`\@=12

\vfill\eject

%
%
\end{document}